%% file: HIN-22-004_temp.tex
\documentclass[11pt,twoside,a4paper,cmspaper,final,collab]{cms-tdr}
\def\svnVersion{1f2bf8e}\def\svnDate{2026/05/29}\def\cmsCernNoTag{CERN-EP-2025-290}\def\cmsCernDate{\today}\def\cmsMessage{Submitted to Physical Review C}
\begin{document}\cmsNoteHeader{HIN-22-004}

\newlength\cmsFigWidth
\ifthenelse{\boolean{cms@external}}{\setlength\cmsFigWidth{0.49\textwidth}}{\setlength\cmsFigWidth{0.65\textwidth}}

\newcommand{\pPb}{\ensuremath{\Pp\text{Pb}}\xspace}

\newcommand{\PomPb}{\text{{~\!I\!P}\text{Pb}}\xspace}
\newcommand{\Pom}{\text{{~\!I\!P}}\xspace}

\newcommand{\pp}{\ensuremath{\Pp\Pp}\xspace}
\newcommand{\PhotonP}{\ensuremath{\Pgg\Pp}\xspace}
\newcommand{\PhotonPb}{\ensuremath{\Pgg\text{Pb}}\xspace}

\newcommand{\DeltaEtaF}{\ensuremath{{\Delta\eta^{\mathrm{F}} }}\xspace}

\newcommand{\Ntrig}{\ensuremath{N_\text{trig}}\xspace}

\newcommand{\Npair}{\ensuremath{N^\text{pair}}\xspace}
\newcommand{\Nassoc}{\ensuremath{N_{\text{assoc}}}\xspace}

\newcommand{\VoneDelta}{\ensuremath{{V_{1\Delta}}}\xspace}
\newcommand{\VtwoDelta}{\ensuremath{{V_{2\Delta}}}\xspace}
\newcommand{\VthreeDelta}{\ensuremath{{V_{3\Delta}}}\xspace}
\newcommand{\VnDelta}{\ensuremath{{V_{n\Delta}}}\xspace}

\newcommand{\vTwo}{\ensuremath{v_2}\xspace}
\newcommand{\vThree}{\ensuremath{v_3}\xspace}

\newcommand{\DeltaEta}{\ensuremath{{\Delta\eta}}\xspace}
\newcommand{\DeltaPhi}{\ensuremath{{\Delta\phi}}\xspace}

\newcommand{\pTassoc}{\ensuremath{\pt^{\text{assoc}}}\xspace}
\newcommand{\pTtrig}{\ensuremath{\pt^{\text{trig}}}\xspace}

\newcommand{\Nch}{\ensuremath{N_{\text{ch}}^{\abs{\eta}<2.4,\,\pt>0.3\GeV}}\xspace}
\newcommand{\Ntrkoffline}{\ensuremath{N_{\text{trk}}^{\text{offline}}}\xspace}
\renewcommand{\arraystretch}{1.15}

\cmsNoteHeader{HIN-22-004}
\title{Dependence of two-particle azimuthal correlations on the forward rapidity gap width in \texorpdfstring{\pPb}{pPb} collisions\\ at \texorpdfstring{$\sqrtsNN = 8.16\TeV$}{sqrt(sNN) = 8.16 TeV}}

\date{\today}

\abstract{
	One of the most striking features of relativistic heavy ion collisions is the presence of collective flow of thousands of produced particles. This flow can be characterized by the Fourier coefficients (\VnDelta) of the azimuthal angular distributions of charged particles, and its existence can be explained by the formation of a quark gluon plasma, which behaves as a fluid. Surprisingly, the angular distributions of particles from very small systems such as proton-lead (\pPb), proton-proton (\pp), electron-positron, and photon-proton (\PhotonP) collisions also exhibit non-zero Fourier coefficients, raising the question of whether collective flow is present. This paper presents measurements of \VnDelta from a sample of \pPb events at $\sqrtsNN=8.16\TeV$ that are enriched in photon-lead (\PhotonPb) and pomeron-lead (\PomPb) interactions by requiring no particles in the proton-going region. Measurements are made as a function of the forward rapidity gap width (the rapidity range in which no particles are found), the transverse momentum of the particles, and the multiplicity of particles in the event. The results are compared to previous measurements of \pp, \pPb, and $\PhotonP+\Pom\Pp$ events as well as modern event generators.
}

\hypersetup{
pdfauthor={CMS Collaboration},
pdftitle={Dependence of two-particle azimuthal correlations on the forward rapidity gap width in pPb collisions at sqrt(sNN) = 8.16 TeV},
pdfsubject={CMS},
pdfkeywords={CMS, pomeron-proton, collectivity}}

\maketitle 

\section{Introduction}

Two-particle azimuthal correlations can provide insights into the so-called collectivity of systems produced in high-energy collisions. Collectivity refers to the correlations among multiple particles in the final state of these interactions. These correlations can help in understanding the system produced in the collision and its hydrodynamic properties. The evidence for collectivity includes both long-range rapidity correlations ($\abs{\DeltaEta}>2$), known as the ``ridge'', for pairs of particles that are close in azimuthal angle ($\phi$), as well as the observation of multiparticle azimuthal correlations in proton-proton (\pp)~\cite{cms:ppfirst,Aad:2015gqa,cms:ppsecond,Khachatryan:2016txc,add_atlas_pp:01}, proton-lead (\pPb)~\cite{add_rhic_pA:01,add_rhic_pA:02,cms:pPbfirst,Aad:2012gla,Aad:2013fja,Abelev:2012ola,Aaij:2015qcq,ABELEV:2013wsa,Khachatryan:2015waa,cms:pPbPbPb_corr_identifiedPar,Aaboud:2017acw,Aaboud:2017blb,CMS:2020qul}, and a wide range of high-energy nucleus-nucleus collisions~\cite{add_rhic_AA:01,add_rhic_AA:04,cms:PbPbfirst,cms:PbPbsecond,star:AAfirst,star:AAsecond,phenix:AAfirst,STAR:oct2019,STAR:oct2018,CMS:2024krd} using the Large Hadron Collider (LHC) at CERN and the Relativistic Heavy Ion Collider (RHIC) at BNL.

Azimuthal correlations between particle pairs can be quantified via their Fourier decomposition, $\rd\Npair / \rd\DeltaPhi \propto 1 + \sum_{n}2\VnDelta \cos(n\DeltaPhi)$, where \VnDelta are the two-particle Fourier coefficients, \Npair is the number of pairs, and \DeltaPhi is the angle difference of the pair. If factorization is assumed, the single-particle azimuthal anisotropy Fourier coefficients $v_n$ can be extracted as $v_n = \sqrt{\VnDelta}$~\cite{Voloshin:1996}. The second- (\vTwo) and third-order (\vThree) coefficients are known as elliptic and triangular flow, respectively~\cite{elliptical_triangular}. In hydrodynamical models, \vTwo and \vThree are directly related to the initial collision geometry and its fluctuations, respectively~\cite{geometry_and_fluctuations_01,geometry_and_fluctuations_02,geometry_and_fluctuations_03}. In general, these Fourier components provide insights into the transport properties of the medium.

In recent years, collective behavior has been found in small systems such as those produced in \pp and \pPb collisions~\cite{cms:ppfirst, cms:pPbfirst}. This has generated a series of studies at RHIC, KEKB, and the LHC~\cite{cms:pPbPbPb_corr_identifiedPar,cms:april2018,Sirunyan:2018nqr,Adam:2015vsf,add_rhic_AA:02,add_rhic_AA:03,Belle:2022fvl}. For high-multiplicity events, the \vTwo and \vThree Fourier coefficients depend upon the hadron species. For \pPb and PbPb collisions, the value of \vTwo scales with the number of valence quarks in the hadron~\cite{cms:pPbPbPb_corr_identifiedPar}. While these results in small systems are consistent with a hydrodynamic expansion, alternative models based on gluon saturation in the initial state can also reproduce the main features of the correlation data (recent reviews are provided, \eg, in Refs.~\cite{Dusling:2015gta, Nagle:2018eea}). While long-range azimuthal correlations have been established in inclusive high-multiplicity \pp and \pPb events, their microscopic origin remains actively debated. They thus reinforce the interest in correlation measurements for other small systems, such as those produced in electron-positron, electron-proton, photon-proton (\PhotonP), and low multiplicity \pp collisions~\cite{alepCorr:2019, Chen:2023njr, Belle:2022fvl, zeus:Dec2019_ep, CMSmurillo, ALICE:2025bwp}, as well as in photon-lead (\PhotonPb)~\cite{Atlas_photon_pb}, oxygen-oxygen, and neon-neon systems~\cite{CMS:2025tga}, and in high-multiplicity jets~\cite{CMS:2023iam}.

Event selections that suppress multiparton interactions and modify color reconnection processes provide an important discriminator between scenarios based on a collective final-state response and those dominated by initial-state correlations. Selecting events with no particles in a portion of the proton-going $\eta$ region (called the ``forward rapidity gap width'') can offer such a handle since this enhances semi-exclusive topologies associated with color-singlet exchange and photo-induced processes. Therefore, these events allow the study of azimuthal correlations in samples that are complementary to the inclusive measurements. The forward rapidity gap width (\DeltaEtaF) is defined as the width of the rapidity region (starting at $\eta=3$ and extending inward in the proton-going direction) where no particles are detected. This paper presents the first search for long-range azimuthal correlations as a function of the forward rapidity gap width in \pPb collisions.

The \pPb collision data were taken with the CMS experiment at the LHC at a center-of-mass energy of $\sqrtsNN=8.16\TeV$~\cite{Chatrchyan:2008zzk}. Events with a large \DeltaEtaF are asymmetric in pseudorapidity, with most activity in the part of the detector opposite to the direction of the outgoing proton beam, and almost no activity on the other side. This asymmetry enhances semi-exclusive events, in which the proton remains intact, or dissociates into a low-mass state that escapes undetected, while the lead nucleus breaks up. The sample contains diffractive events, where the \pPb interaction is mediated by an object with the vacuum quantum numbers, the so-called pomeron (\Pom)~\cite{Regge:1959mz}; smaller contributions from \PhotonPb and nondiffractive interactions is also present.

The sample analyzed here was also studied in Ref.~\cite{HIN1819}, where the differential cross section as a function of \DeltaEtaF for \pPb events was measured. Two types of events were selected: those in which the final-state particles mostly go in the direction of the outgoing proton, and those in which most of the activity is in the direction of the outgoing lead nucleus. The former sample was used to measure two-particle correlations in \PhotonP events~\cite{CMSmurillo}. The latter is analyzed in this paper. This sample has a varying mix of diffractive, nondiffractive and \PhotonPb events with different levels of contributions across \DeltaEtaF. The exact level of these contributions is still an open topic since the yield obtained in Ref.~\cite{HIN1819} is not completely accounted for by current theoretical models. For large \DeltaEtaF $(>2.5)$, this sample consists of diffractive events $(\sim43.5\%)$ with a small contribution of nondiffractive events $(\sim5\%)$, according to the \textsc{epos-lhc} Monte Carlo (MC) generator~\cite{EPOS}. The rest of the yield, including \PhotonPb events, is not described by the MC models. For this ``diffraction-enhanced'' region with $\DeltaEtaF>2.5$, two-particle correlations are investigated in detail, thus extending the cross section measurements presented in Ref.~\cite{HIN1819}. Tabulated results are provided in the HEPData record for this analysis~\cite{hepdata}.

\begin{figure*}[ht]
 \centering
  \includegraphics[width=0.3\textwidth]{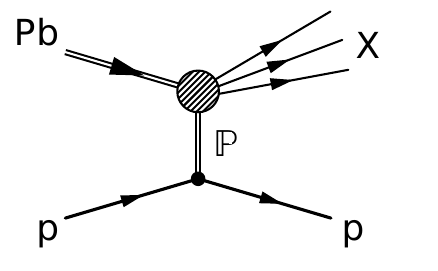}
  \includegraphics[width=0.4\textwidth]{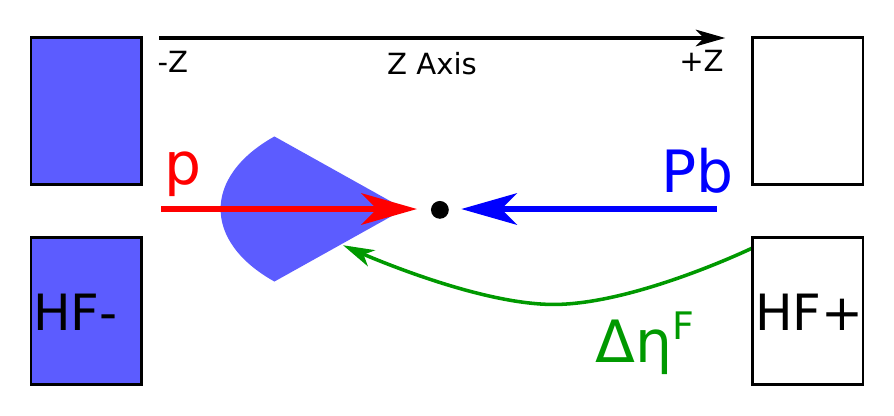}
 \caption{Sketch of a single-diffractive \pPb event (left). The interaction proceeds via the exchange of a pomeron (\Pom); the proton remains intact or dissociates into a low-mass state, and the nucleus breaks up (X). A single-diffractive event as seen in the detector (right). Activity is observed only in one side of the detector and a large forward rapidity gap (\DeltaEtaF) is present. The blue cone represents the outgoing particles from the interaction, while the blue shading indicates the detected activity. Both diagrams are extracted from Ref.~\cite{HIN1819}.}
	\label{Fig:diagrama_proceso}
\end{figure*}

In the present analysis, the selection was designed to enhance the single-diffractive \PomPb contribution (Fig.~\ref{Fig:diagrama_proceso}, left). In this process, the proton emits a pomeron that interacts with the lead nucleus, causing the breakup of the nucleus, while the proton remains intact, or dissociates into a low-mass state that escapes undetected. The results of this study will serve as a reference for predictions from \Pom-based models, such as those following the Ingelman--Schlein approach~\cite{Ingelman:1984ns}, which assign a pomeron flux and structure to the proton and employ tools like \textsc{Angantyr}~\cite{Bierlich:2018xfw} to model the pomeron-nucleus interaction.

Both single-diffractive and \PhotonPb events are characterized by rapidity gaps, as shown in Fig.~\ref{Fig:diagrama_proceso} (right), where the activity due to the breakup of the lead nucleus is indicated in blue.

\section{Experimental apparatus}
 
The central feature of the CMS apparatus is a superconducting solenoid of 6 m internal diameter, providing a magnetic field of 3.8 T. Within the solenoid volume is the silicon tracker, a lead tungstate crystal electromagnetic calorimeter, and a brass and scintillator hadron calorimeter, each composed of a barrel (EB and HB) and two endcap (EE and HE) sections, and covering the pseudorapidity range $\abs{\eta} < 3.0$. The silicon tracker measures charged particles within the range $\abs{\eta}< 2.5$; it consists of 1\,440 silicon pixel and 15\,148 silicon strip detector modules and provides an impact parameter resolution of $\approx15\mum$. The event selection for this analysis exploits the HF detectors ($\text{HF}+$ and $\text{HF}-$) in the $3.0<\abs{\eta} < 5.2$ region, which use quartz fibers embedded in a steel absorber~\cite{CMS:2012tda}. 

The analysis in the region $\abs{\eta} < 3.0$ is based upon objects produced by the CMS particle-flow (PF) algorithm~\cite{Sirunyan:2017ulk}, which reconstructs and identifies final-state particles with an optimized combination of information from the various elements of the CMS detector. 

A more detailed description of the CMS detector, together with a definition of the relevant kinematic variables, can be found in Ref.~\cite{Chatrchyan:2008zzk}. The coordinate system used in this paper is defined as follows: $x$ points in the direction of the center of the LHC ring, $y$ is upwards and $z$ is along the beam line according to the right-hand rule. The $\eta$ is defined such as to have positive values when $z>0$.

\section{Data samples}

The analysis is performed using data recorded by CMS during the LHC \pPb run in 2016 corresponding to an integrated luminosity of 174.5\nbinv~\cite{CMS:2018fkg}. The beam energies were 6.5\TeV for the protons and 2.56\TeV per nucleon for the lead nuclei in the laboratory frame, resulting in a center-of-mass energy per nucleon pair (\sqrtsNN) of 8.16\TeV. The data include samples with the proton going towards the positive and towards the negative $\eta$ direction. As a result of the energy difference between the colliding beams, the proton-nucleus center-of-mass frame for a \pPb collision is not at rest with respect to the laboratory frame. Massless particles emitted at $\eta_{\text{cm}} = 0$ in the \pPb center-of-mass frame are detected at $\eta = +0.465$ in the laboratory frame.

The data were selected online with a two level trigger system consisting of the level-1, based on custom hardware processors, and the high-level trigger (HLT), running a version of the full event reconstruction software optimized for fast processing~\cite{Sirunyan:2017uyl, Khachatryan:2016bia}.

Monte Carlo samples are used for comparing models to the measurements. Samples are generated with \textsc{epos-lhc}~\cite{EPOS}, and \PYTHIA 8.3.10 event generators~\cite{Bierlich:2022pfr}. The \textsc{epos-lhc} generator is based on the Gribov--Regge model~\cite{EPOS}, which includes nondiffractive (ND), central-diffractive (CD), single-diffractive (SD), and double-diffractive (DD) events, but no photo-induced processes; collective effects are simulated. The \textsc{epos-lhc} samples are subsequently processed with \GEANTfour to simulate the response of the CMS detector~\cite{Agostinelli:2002hh}. Samples were produced with the proton beam in both directions. The \PYTHIA 8.3.10 generator is based on the \textsc{Angantyr} model~\cite{Bierlich:2018xfw}. Separate samples of ND, SD, DD events were produced. The generator does not include collective effects nor photo-induced processes. The \PYTHIA 8.3.10 samples were only used to compare to data corrected to the generator level and, therefore, were not processed to detector level using \GEANTfour.

\section{Event selection}

The event selection and the definition of the data samples follow closely those of Ref.~\cite{HIN1819}. Events are selected by the level-1 trigger if energy deposits are present in at least one of the HF calorimeters above a threshold of approximately 1\GeV. The presence of at least one track with $\pt>0.4\GeV$ in the pixel tracker is required at the HLT. The track reconstruction is performed online as part of the HLT trigger with reconstruction algorithms identical to those used offline~\cite{highqualitytracks}. For each event, the reconstructed vertex with the highest number of associated tracks is selected as the primary vertex (PV).
 
The reconstructed PV is required to be within 15\cm of the nominal interaction point along the beam axis $(z)$ and within 0.15\cm in the transverse plane. High-purity tracks~\cite{highqualitytracks} are used. At least two reconstructed tracks with $\abs{\eta}<2.4$, $\pt>0.4\GeV$ are required to be associated with the PV. Beam related background is suppressed by rejecting events for which less than 25\% of all reconstructed tracks pass the track selection criteria mentioned above. Additionally, the impact parameter significance of reconstructed tracks with respect to the PV in the longitudinal and transverse directions is required to be less than 3 standard deviations. Finally, the relative uncertainty of the track \pt is required to be less than 10\%.

A forward rapidity gap is a region with low detector activity and is defined as follows. The region $-3.0<\eta<3.0$ is divided into 12 bins that are 0.5 units wide. Energy thresholds are assigned to each bin such that at least 99.7\% of the signals due to detector noise or beam gas events~\cite{HIN1819} are below the threshold. These are 6\GeV for the central region $\abs{\eta}<2.5$, for which all PF candidates are considered, and 13.4\GeV for $\abs{\eta}>2.5$ for which only neutral hadronic PF candidates are considered due to noise in the electromagnetic calorimeter. A bin is considered empty if the total energy from all PF candidates is below the threshold. In addition, for the central region $\abs{\eta}<2.5$, no high-purity tracks with $\pt> 200\MeV$ are allowed. The \DeltaEtaF variable is then defined as the difference between $\eta = 3.0$ and the higher edge of the first nonempty $\eta$ bin. The definition is reversed when the direction of the proton beam is towards negative values of $\eta$.

Events are selected as follows. The hadronic forward $\text{HF}+$ calorimeter on the positive $\eta$ side is required to have all towers with an energy deposit below 2\GeV. The $\text{HF}-$ calorimeter on the negative side is required to have at least one tower with an energy deposit above 10\GeV. Additionally, the $\eta$ bin $[+2.5, +3.0)$ is required to be empty, reflecting the lack of activity in the direction of the outgoing proton. Events with a presence of a high-purity track with $\eta>2.5$ and $\pt>200\MeV$ were removed since their effect on \DeltaEtaF is not well understood. This selection is referred to in the following as the ``nominal selection'' and the resulting sample as the ``nominal sample''. The definitions are reversed in $\eta$ when the direction of the proton beam is towards negative values of $\eta$. The analysis is performed in different \DeltaEtaF bins from 0.5 to 2.5 in intervals 0.5 units wide and a final bin from 2.5 onwards.

In Ref.~\cite{HIN1819} it was found that the \textsc{epos-lhc} generator predicts a \DeltaEtaF spectrum with a shape reasonably consistent with that of the data but with a different yield. The results suggest only a small nondiffractive contribution when the rapidity gap cut $\DeltaEtaF > 2.5$ is applied; an additional set of measurements with such a cut is carried out here. This selection is referred to as ``diffraction-enhanced selection'' in the following, and the resulting sample is obtained by applying the nominal selection but restricting the events to the last \DeltaEtaF bin. For events with rapidity gaps on the negative side (not analyzed here), MC studies indicate that they are dominated by \PhotonP interactions~\cite{HIN1819}.

\begin{figure*}[ht!]
\centering
\includegraphics[width=0.75\textwidth]{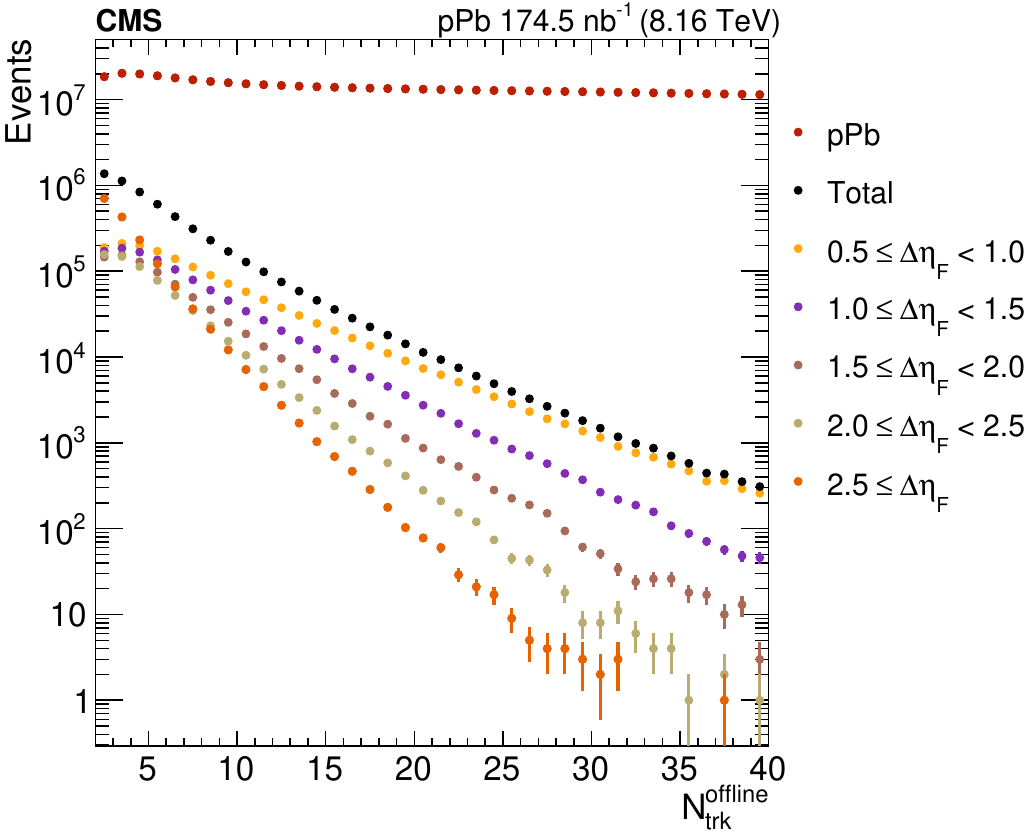}
\caption{The \Ntrkoffline spectra for inclusive \pPb events (labeled ``\pPb'', from Ref.~\cite{CMSmurillo}) and events from the nominal sample in different \DeltaEtaF bins. The data points are located in the center of the bin of their corresponding \Ntrkoffline category. The vertical bars indicate the size of the statistical uncertainties.}
\label{fig:spectraDeltaEta}
\end{figure*} 

\begin{figure}[ht!]
\centering
\includegraphics[width=\cmsFigWidth]{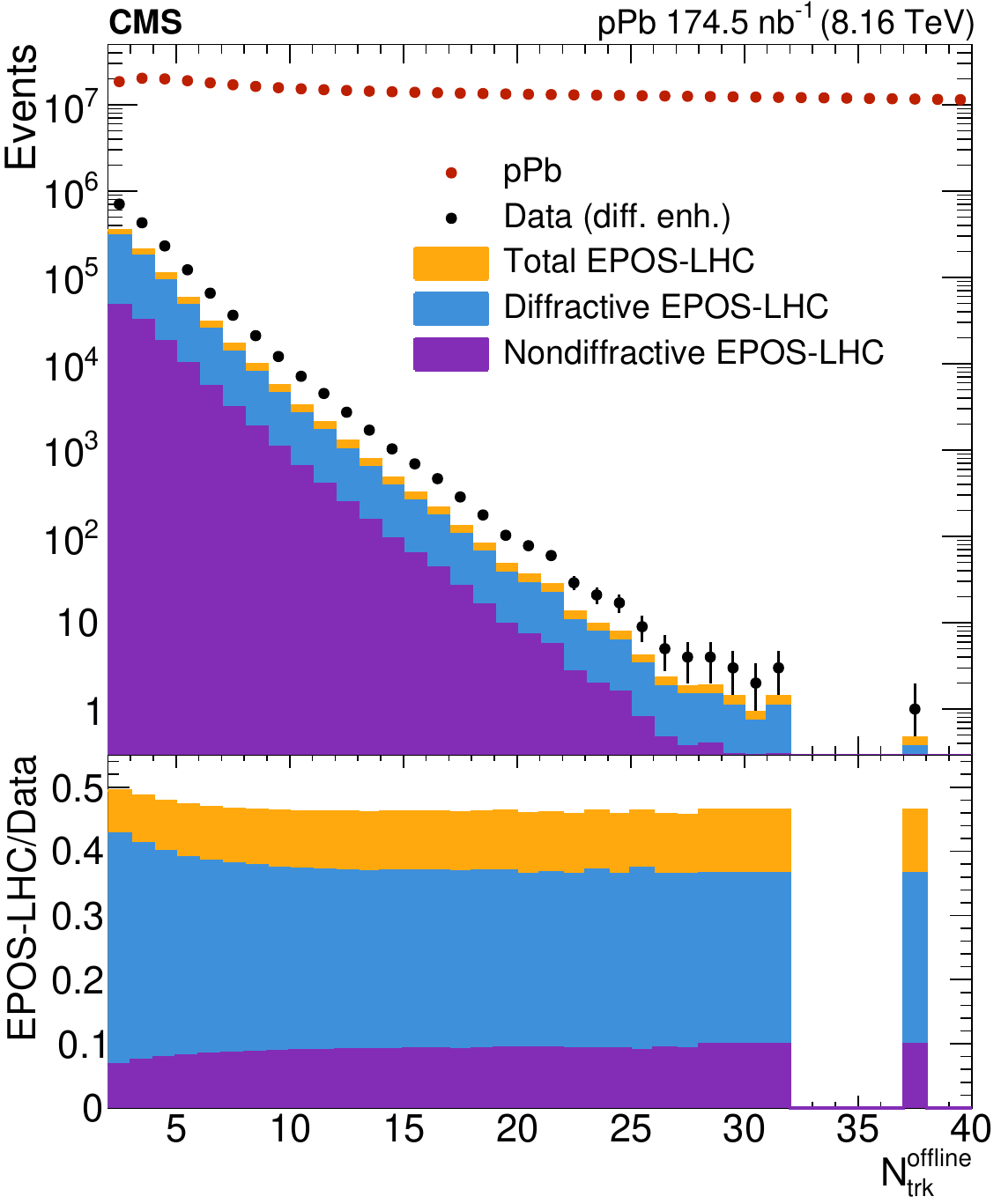}
\caption{The \Ntrkoffline spectra for inclusive \pPb events (labeled ``\pPb'', from Ref.~\cite{CMSmurillo}) and inclusive diffraction-enhanced events (labeled ``Data (diff. enh.)''). The distribution for the inclusive diffraction-enhanced events is the same as that shown in orange in Fig.~\ref{fig:spectraDeltaEta} ($\DeltaEtaF > 2.5$). Predictions based on weighting the data by the fractional cross sections from \textsc{epos-lhc} are also shown. The lower panel shows the ratio of the predictions to the data. The vertical bars indicate the size of the statistical uncertainties.}
\label{fig:spectra}
\end{figure} 

Track multiplicity distributions, \Ntrkoffline, are obtained by counting the tracks satisfying the requirements listed in Section 4 and with $\pt>0.4\GeV$. In Figs.~\ref{fig:spectraDeltaEta} and~\ref{fig:spectra}, the resulting \Ntrkoffline distributions for the data are shown for the nominal and the inclusive diffraction-enhanced selection, respectively. The inclusive distribution in \pPb collisions, from Ref.~\cite{CMSmurillo}, is also shown in both figures. In Fig.~\ref{fig:spectraDeltaEta}, the distributions for different bins of \DeltaEtaF with the nominal selection are shown along with the inclusive \pPb data. The multiplicity of the diffractive samples is on average small, and the distributions are steeper than for the inclusive sample. Furthermore, the slopes increase with \DeltaEtaF. The mean track multiplicity $\langle \Ntrkoffline \rangle$ decreases by 48\% when moving from $0.5<\DeltaEtaF<1.0$ to $\DeltaEtaF>2.5$. The most populated bin is the one with \DeltaEtaF in the range $[0.5,1)$, whereas the rest of the \DeltaEtaF bins have multiplicities $\Ntrkoffline<50$, comparable to the \PhotonP case~\cite{CMSmurillo}.

In Fig.~\ref{fig:spectra}, the distribution of \Ntrkoffline for the inclusive diffraction-enhanced selection is shown, along with MC predictions from Ref.~\cite{HIN1819}. The inclusive \pPb points~\cite{CMSmurillo} are the same as in Fig.~\ref{fig:spectraDeltaEta}. The MC predictions are obtained by weighting the data distribution with the cross sections determined in Ref.~\cite{HIN1819} from the \textsc{epos-lhc} generator for the \DeltaEtaF value of each event. The predictions for the diffractive and nondiffractive contributions are shown separately. The \PhotonPb contribution is not included. The \textsc{epos-lhc} model predicts a distribution with very similar shape to that of the data, as can be seen from the ratio of MC over data yields, which is mostly flat.

For the diffraction-enhanced selection, the analysis is performed in the following three bins of \Ntrkoffline: $[2,5)$, $[5,7)$, and $[7,40)$. For the \textsc{epos-lhc} samples, the higher \Ntrkoffline bins are combined in order to improve the statistical accuracy, so the bins are $[2,5)$ and $[5,40)$.

\section{Analysis technique}

To ensure high tracking efficiency, only tracks within $\abs{\eta} < 2.4$ and $0.3<\pt<3\GeV$ are used in the analysis. The \Ntrkoffline is taken as the number of tracks from the PV with $\abs{\eta} < 2.4$ and $\pt>0.4\GeV$. The analysis techniques for two-particle correlations are identical to those used in past CMS measurements~\cite{cms:ppsecond,cms:pPbfirst,cms:PbPbsecond,CMSmurillo}, and are described below. Because of the limited \Ntrkoffline range for the selected events, no low-multiplicity subtraction technique is implemented~\cite{Sirunyan:2017uyl}.

For each bin in the nominal and diffraction-enhanced selection, tracks tagged as ``trigger particles'' are those whose \pt, referred to as \pTtrig, is within a given range (discussed below). The number of trigger particles in the event is denoted by \Ntrig. Particle pairs are formed by associating each trigger particle with the remaining charged tracks within a specified $\pt\equiv \pTassoc$ interval. For the present analysis, identical ranges are used for \pTtrig and \pTassoc. Two different \pt ranges are studied, $0.3<\pt<3$ and $1<\pt<3\GeV$, identical to the ones used in previous studies of the ridge~\cite{cms:pPbfirst} and of correlated azimuthal anisotropies~\cite{Sirunyan:2017uyl} in \pPb collisions. For $1<\pt<3\GeV$, only the bin $2\leq \Ntrkoffline<40$ is studied.

The two-dimensional (2D) correlation function is defined as
\begin{equation}
\frac{1}{\Ntrig}\frac{\rd^2\Npair}{\rd\DeltaEta \rd\DeltaPhi} = B(0,0)\frac{S(\DeltaEta, \DeltaPhi)}{B(\DeltaEta, \DeltaPhi)}, 
\end{equation}
where \DeltaEta and \DeltaPhi are the differences between the $\eta$ and $\phi$ values of the trigger particle and the other track of the pair. The quantity \Npair is the number of pairs. The same-event pair distribution, $S(\DeltaEta, \DeltaPhi)$, represents the yield of particle pairs from the event under consideration. The mixed-event pair distribution $B(\DeltaEta, \DeltaPhi)$ is constructed by pairing the trigger particles in each event with the associated charged particles from 100 different randomly selected events in the same 0.5\cm wide $z$ vertex range and from the same track multiplicity class. The same-event and mixed-event pair distributions are first calculated for each event, and then averaged over all the events within the track multiplicity class. The mixed-event distribution is normalized to the sum of background events. The ratio $B(0,0)/B(\DeltaEta, \DeltaPhi)$ accounts for pair acceptance effects, with $B(0,0)$ representing the mixed-event associated yield for both particles of the pair going in the same direction and thus having maximum pair acceptance. 

A correction for tracking inefficiency is applied by weighting each charged particle by the inverse of the track reconstruction efficiency obtained in Ref.~\cite{CMS:2010mua}, so that these distributions can be compared with generator-level simulations. This correction is applied to the \Ntrkoffline distribution in the diffraction-enhanced sample. The new corrected quantity is called \Nch and the average within a \Ntrkoffline bin is used to present the final results for the diffraction-enhanced sample.

The charged particle azimuthal anisotropy harmonics can be extracted via a Fourier decomposition of long-range two-particle \DeltaPhi correlation functions, obtained by averaging the 2D two-particle correlation function over $\abs{\DeltaEta}>2$ (to remove short-range correlations, such as those from jet fragmentation~\cite{ALICE:2016wkc}). The fitted functional form is:
\begin{equation}
\frac{1}{\Ntrig}\frac{\rd \Npair}{\rd\DeltaPhi} = \frac{\Nassoc}{2\pi}[1 + \sum_{n}2\VnDelta\cos(n\DeltaPhi)],
\end{equation}
where \VnDelta are the Fourier coefficients and \Nassoc represents the average number of pairs per trigger particle for a given (\pTtrig, \pTassoc) bin. In this analysis, we included terms up to $n=3$, because including higher order \VnDelta terms has a negligible effect on the results of the Fourier fit.

\begin{figure*}[!htb]
\label{an:05}
 \centering
  	\includegraphics[width=0.49\textwidth]{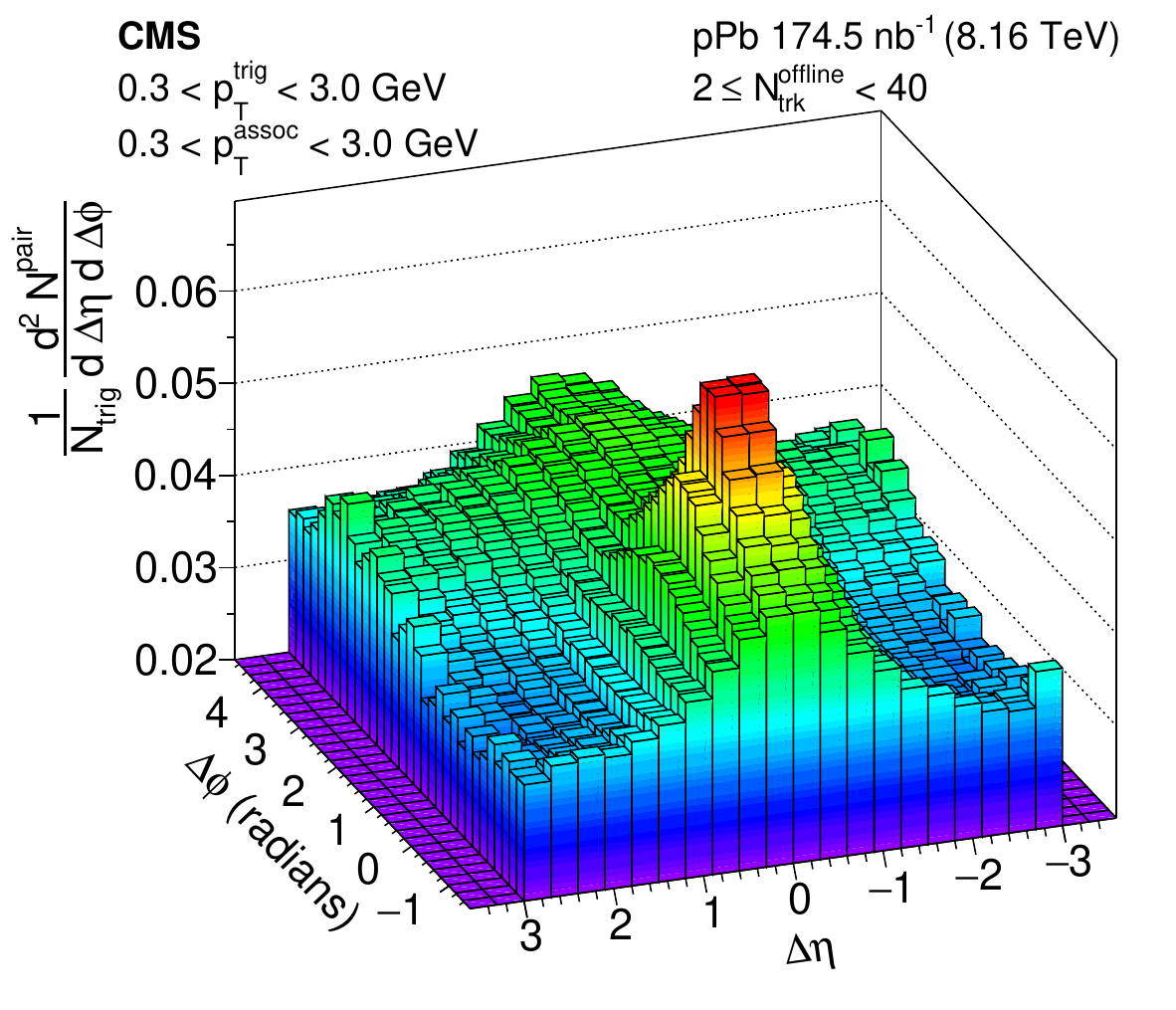}
	\includegraphics[width=0.49\textwidth]{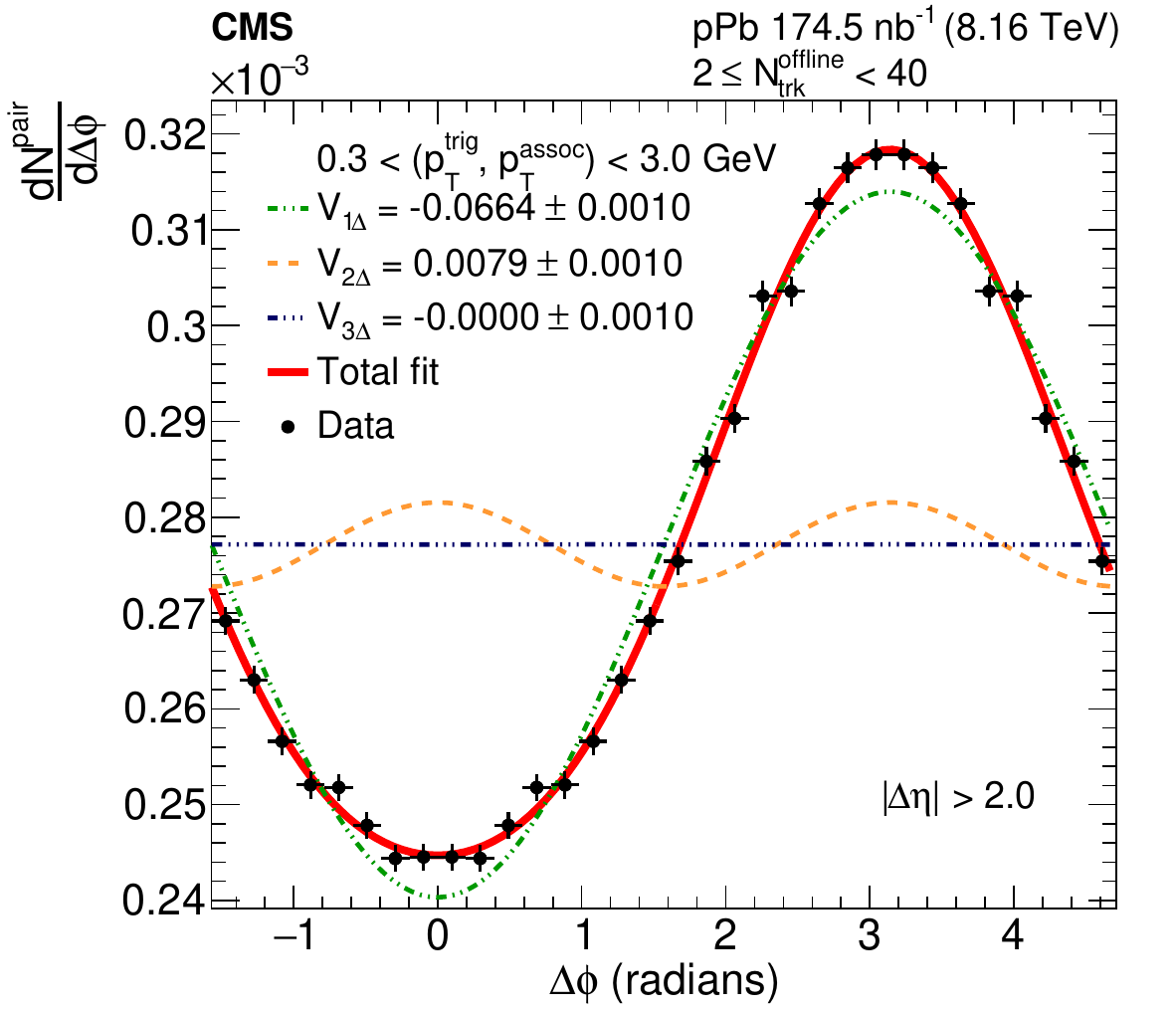}
 \caption{The 2D correlation distribution for multiplicity (left) and its projection on the \DeltaPhi axis (right) for $\abs{\DeltaEta}>2$, $2\leq\Ntrkoffline<40$, and $0.3<\pTtrig<3.0\GeV$. The filled circles indicate the data and the vertical bars indicate the size of the statistical uncertainties. The one-dimensional \DeltaPhi distributions are symmetrized by construction around $\DeltaPhi = 0$ and $\DeltaPhi = \pi$, so all the data are contained in $[0, \pi]$ and are averaged over $\abs{\DeltaEta}>2$. The plotted range in \DeltaPhi extends to $[-\pi, 3\pi/2]$ for visualization purposes. The Fourier coefficients \VnDelta are fitted over the \DeltaPhi range $[0, \pi]$. The dashed lines show the resulting components \VoneDelta, \VtwoDelta, and \VthreeDelta, along with the fitted values and statistical uncertainties. }
  \label{fig:CorrelationPlots}
\end{figure*}

Figure~\ref{fig:CorrelationPlots} (left) shows the two-particle correlation function for the events from the diffractive-enhanced sample in the range of $2 \leq \Ntrkoffline<40$ as a function of \DeltaEta and \DeltaPhi, for $0.3<\pTtrig<3.0\GeV$. The distribution exhibits a large jet peak centered at $\DeltaEta = \DeltaPhi = 0$ as well as a broader structure from the recoiling jets centered at $\DeltaEta = 0$ and $\DeltaPhi = \pi$. A ``ridge''-like structure at $\abs{\DeltaPhi}\approx0$ for $\abs{\DeltaEta}>2$ is not visible. Figure~\ref{fig:CorrelationPlots} (right) shows the projection of the 2D correlation function on the \DeltaPhi axis for $\abs{\DeltaEta}>2$, \ie, away from the jet fragmentation peak. Fits to a Fourier series performed over the \DeltaPhi range $[0, \pi]$, which are discussed in Section~\ref{results}, are also shown. 

\section{Systematic uncertainties}

In this section, the sources of the systematic uncertainties in the \VnDelta Fourier coefficients are discussed. The systematic uncertainties are evaluated as a function of \DeltaEtaF and \Nch by varying the way the coefficients are determined.

The general procedure for determining the systematic uncertainties is described in Ref.~\cite{Barlow:2002yb}. The systematic uncertainty is considered to be significant (and taken into account) when it affects the result by more than its statistical uncertainty, having considered the correlation between the sample modified by the systematic variation and the nominal sample.

The systematic effects considered are the following:

\begin{itemize}
   \item The effect of tracking inefficiency and misreconstructed track rate is studied by varying the track quality requirements. The selection thresholds on the significance of the transverse and longitudinal track impact parameters are varied from 2 to 5 standard deviations. In addition, the relative track \pt uncertainty is varied from 5--10\%.
   \item The sensitivity to the PV position along the beam axis $(z_{\text{vtx}})$ is studied by comparing events with different $z_{\text{vtx}}$ locations from $-15$ to $+15\cm$.
   \item The possible contamination from overlapping \pPb collisions in a single event, referred to as pileup, is studied by analyzing two subsets of data with high and low pileup, respectively, corresponding to high and low luminosity periods. On average the high-pileup sample has a pileup of 0.25 while the low-pileup has 0.10.
   \item The trigger efficiency decreases substantially for $\Ntrkoffline < 10$. To study the effect of this decrease, a parallel sample is studied with events weighted by the inverse of the efficiency, which depends on \Ntrkoffline. The full difference of the results with and without the reweighting is taken as the uncertainty.
   \item When the proton beam direction changes, the detectors used for the measurement are different. To study this effect, the results obtained for the two proton beam directions are compared.
\end{itemize}

For \VtwoDelta, the pileup and the change of beam direction are the largest contributors to the systematic uncertainty for the nominal and diffraction-enhanced samples. For \VoneDelta, the pileup and trigger efficiency are the main contributors. This can be observed in Table~\ref{Tab:syst}, which shows the unweighted average (across all \DeltaEtaF bins) of the systematic uncertainty contribution by source for the nominal selection. The individual systematic uncertainties are added in quadrature and amount to about twice the statistical uncertainty. The total uncertainty is obtained by summing the statistical and systematic uncertainties in quadrature. The resulting uncertainties for the diffraction-enhanced sample are of the same order of magnitude as those obtained in \PhotonP interactions~\cite{CMSmurillo} for $\Nch<5$.

\begin{table}[htb!]
	\centering
	\topcaption[Unweighted average systematic uncertainty contribution by source for the nominal selection.]{Unweighted average systematic uncertainty contribution by source for the nominal selection.}
	\begin{scotch}{l c c c}
		\multicolumn{3}{c}{Average systematic uncertainty contribution by source} \\ 
		\hline
		&$\VoneDelta(\times10^{3})$ &$\VtwoDelta(\times10^{3})$ \\
		
		\hline
		Track quality & 0.10 & 0.12 \\
		Pileup & 1.1 & 1.5 \\
		Beam direction & 0.35 & 0.73 \\
		PV position & 0.53 & 0.25 \\
		Trigger efficiency & 0.96 & 0.13 \\
	\end{scotch}
	\label{Tab:syst}
\end{table}

\section{Results}\label{results}

\begin{figure*}[!ht]
	\centering
	\includegraphics[width=0.49\textwidth]{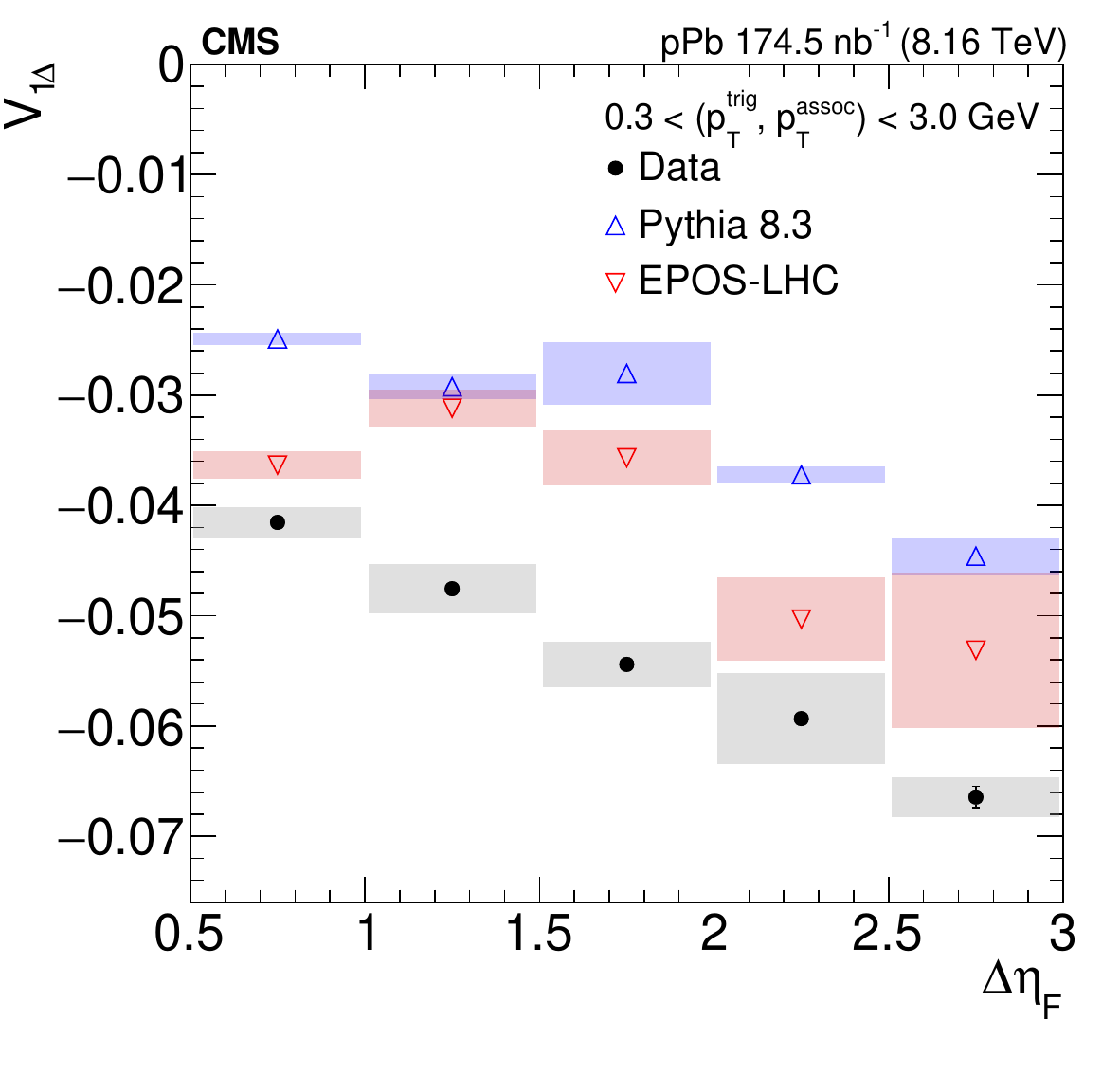}
	\includegraphics[width=0.49\textwidth]{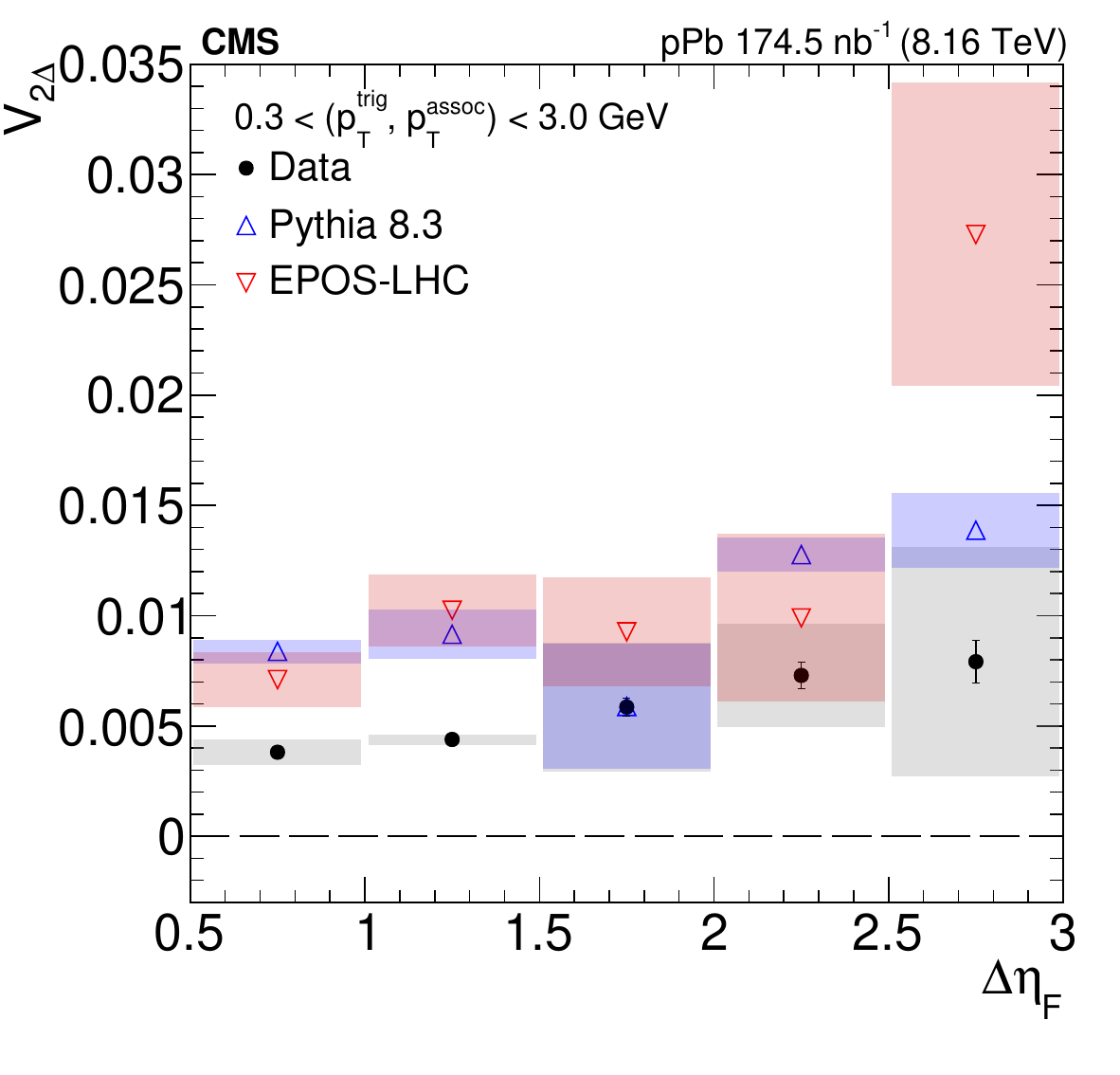}
	\caption{Summary of measurements of \VoneDelta (left) and \VtwoDelta (right) as a function of \DeltaEtaF. The results are for tracks with $0.3<\pTtrig<3.0\GeV$. The filled circles indicate the data. The open triangles indicate the simulation, in blue for \PYTHIA 8.3.10 and red inverse-triangles for \textsc{epos-lhc}. The data points are plotted at the center of the corresponding \DeltaEtaF bin. The height of the shaded bands indicates the systematic and statistical uncertainties for data and simulation, respectively. The vertical bars indicate the size of the statistical uncertainties for data. The width of the shaded bands indicate the size of the \DeltaEtaF bin.
	}
	\label{fig:resultados_frg}
\end{figure*}

The results for \VoneDelta and \VtwoDelta as a function of \DeltaEtaF are shown in Fig.~\ref{fig:resultados_frg} and listed in Table~\ref{Tab:resultados0}. A result is considered consistent with another data point or zero if the comparison meets the 95\% confidence level (\CL) criterion. The \VoneDelta coefficients are negative and decrease with \DeltaEtaF. The central values of the \VtwoDelta coefficients increase monotonically with \DeltaEtaF, although they are consistent with zero within uncertainties in two of the bins. The \PYTHIA 8.3.10 predictions for \VoneDelta are higher than the data and decrease with \DeltaEtaF. Comparing each \DeltaEtaF bin, the \textsc{epos-lhc} predictions for \VoneDelta are consistent with the data except for the bins $[1,1.5)$ and $[1.5,2)$ where they are higher than the data. The \VtwoDelta predictions for both simulations are consistent bin-by-bin with the data except for bins $[0.5,1)$ (in \PYTHIA 8.3.10), and $[1,1.5)$ (for both simulations), where the predictions are higher than the data.

\begin{figure*}[!ht]
	\centering
	\includegraphics[width=0.49\textwidth]{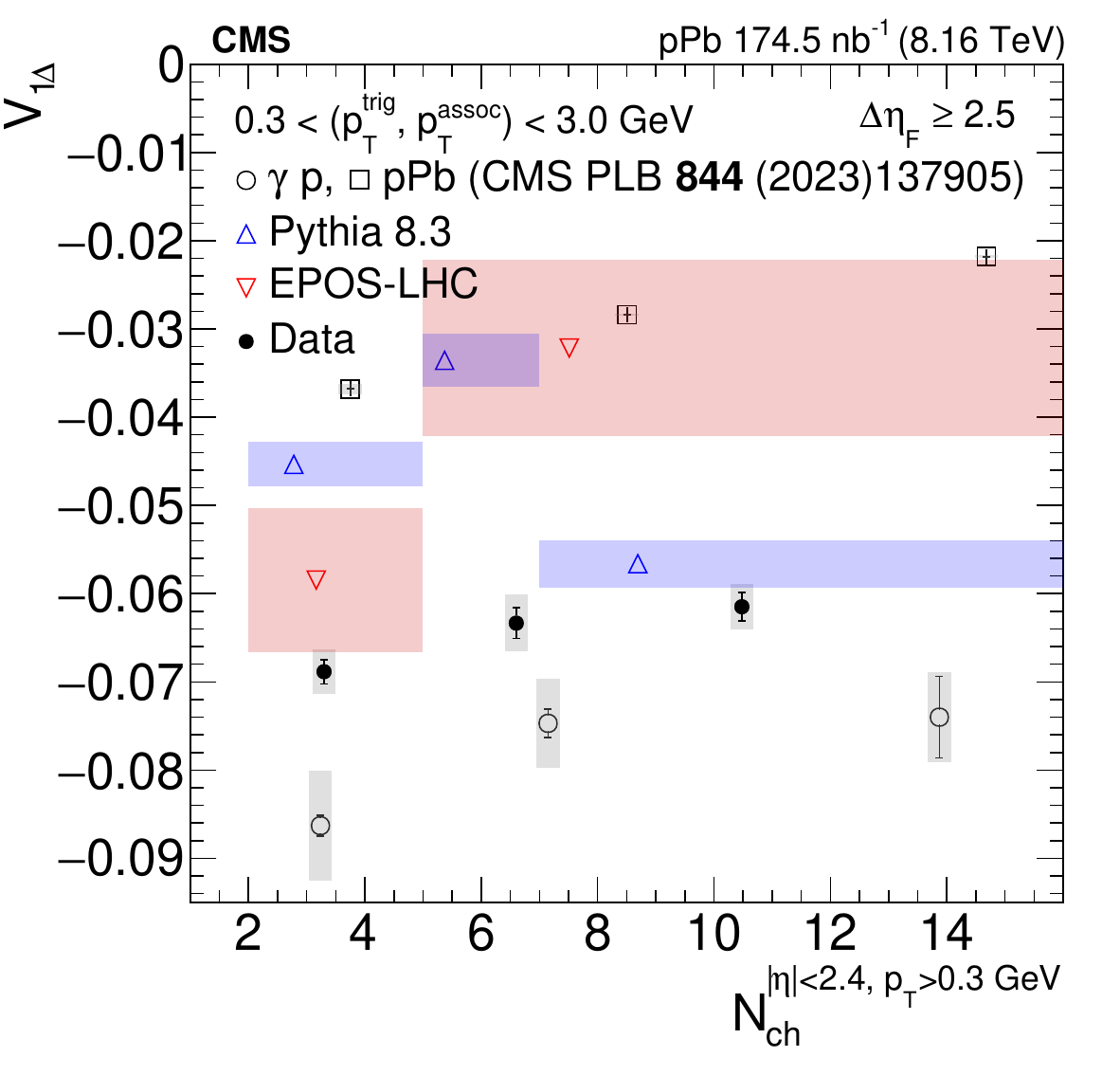}
	\includegraphics[width=0.49\textwidth]{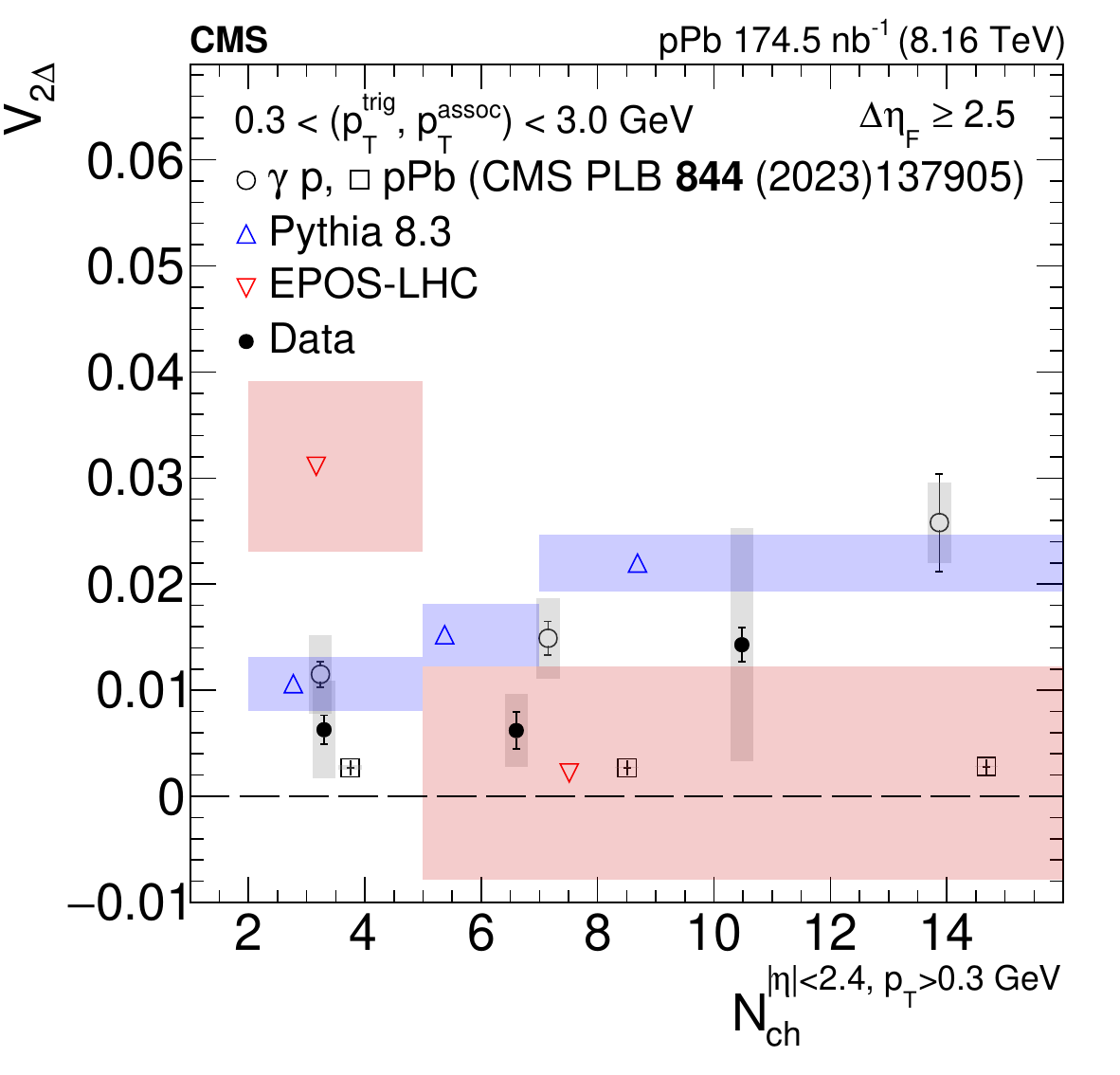}
	\caption{Summary of measurements of \VoneDelta (left) and \VtwoDelta (right) for the diffraction-enhanced events for different classes of \Nch. The results are for tracks with $0.3<\pTtrig<3.0\GeV$. The filled circles indicate the data. The open triangles indicate the simulation, in blue for \PYTHIA 8.3.10 and red inverse triangles for \textsc{epos-lhc}. The data points are plotted at the mean of \Nch within the corresponding \Ntrkoffline bin. The height of the shaded bands indicates the systematic and statistical uncertainties for data and simulation, respectively. The vertical bars indicate the size of the statistical uncertainties for data. The inclusive \pPb and \PhotonP results from Ref.~\cite{CMSmurillo} are also shown as open circles and open squares, respectively.}
	\label{fig:resultados_ntrk}
\end{figure*}

The results for \VoneDelta and \VtwoDelta as a function of \Nch for the diffraction-enhanced sample are shown in Fig.~\ref{fig:resultados_ntrk} and listed in Table~\ref{Tab:resultados1}. The inclusive \pPb and \PhotonP data from Ref.~\cite{CMSmurillo} are also shown, along with the \PYTHIA 8.3.10 and \textsc{epos-lhc} model predictions. The measured values of \VoneDelta consistently fall below the \pPb results in all \Nch ranges, but are in qualitative agreement with the \PhotonP data. For \VtwoDelta, the present data agrees with both \pPb and \PhotonP results within similar \Nch ranges. Both \textsc{epos-lhc} and \PYTHIA 8.3.10 predictions for \VtwoDelta are in qualitative agreement with the data as a function of multiplicity; however, they tend to overestimate the results for \VoneDelta.

The measured values of \vTwo are shown in the last column of Table~\ref{Tab:resultados0} for the \DeltaEtaF bins $[0.5,1)$, $[1,1.5)$, and $[2,2.5)$. The uncertainties in \vTwo were propagated from \VtwoDelta using MC pseudo-experiments. In the rest of the \DeltaEtaF bins, as well as in all the \Nch regions for the diffraction-enhanced sample, the data are consistent with zero and 95\% \CL upper limits are derived as shown in the last column of Tables~\ref{Tab:resultados0} and~\ref{Tab:resultados1} (denoted by the $<$ symbol). The \vTwo limits remain approximately constant across \DeltaEtaF bins in the nominal selection. The largest value is 0.13, which corresponds to the largest \DeltaEtaF bin (same as the inclusive bin in the diffraction-enhanced selection). The limits remain relatively stable across \Nch regions for $0.3<\pTtrig<3.0\GeV$. Table~\ref{Tab:resultados1} (lower) shows the results for \vTwo in \PhotonP events~\cite{CMSmurillo}. The 95\% \CL upper limits on \vTwo are similar to the ones in \PhotonP data, which fall between 0.1--0.2 for comparable \Nch and \pTtrig ranges, with no evidence for the presence of a significant ridge-like structure.

\begin{table*}[htb!]
	\centering
	\topcaption[Measured values of \VnDelta as a function of \DeltaEtaF. The uncertainties include both statistical and systematic components. The table shows the \vTwo results when they are nonzero. For the other cases, upper limits at 95\% \CL are displayed with the $<$ symbol.]{Measured values of \VnDelta as a function of \DeltaEtaF. The uncertainties include both statistical and systematic components. The table shows the \vTwo results when they are nonzero. For the other cases, upper limits at 95\% \CL are displayed with the $<$ symbol.}
	\begin{scotch}{l c c c c}
		\multicolumn{4}{c}{$0.3<\pTtrig<3.0\GeV$} \\ 
		\hline
		&$\VoneDelta(\times10^{3})$&$\VtwoDelta(\times10^{3})$&\vTwo \\
		\hline
		$0.5 \leq \DeltaEtaF < 1.0$ & $-41.5\pm1.4$ & $3.81\pm0.61$ & $0.0617\pm0.0099$ \\
		$1.0 \leq \DeltaEtaF < 1.5$ & $-47.6\pm2.2$ & $4.39\pm0.36$ & $0.0663\pm0.0054$ \\
		$1.5 \leq \DeltaEtaF < 2.0$ & $-54.4\pm2.1$ & $5.9\pm3.0$ & $<0.10$ \\
		$2.0 \leq \DeltaEtaF < 2.5$ & $-59.3\pm4.1$ & $7.3\pm2.4$ & $0.085\pm0.028$ \\
		$\DeltaEtaF \geq 2.5$ & $-66.5\pm2.1$ & $7.9\pm5.3$ & $<0.13$ \\
	\end{scotch}
	\label{Tab:resultados0}
\end{table*}

\begin{table*}[htb!]
	\centering
	\topcaption[Measured values of \VnDelta for the diffractive-enhanced sample as a function of $\langle \Nch \rangle$ and \Ntrkoffline for both \pTtrig bins. The uncertainties include both statistical and systematic components. The table also shows the 95\% \CL upper limits on \vTwo displayed with the $<$ symbol. In the lower table, \vTwo values for \PhotonP events for $0.3<\pTtrig<3.0\GeV$~\cite{CMSmurillo} in similar \Nch ranges are given.]{Measured values of \VnDelta for the diffractive-enhanced sample as a function of $\langle \Nch \rangle$ and \Ntrkoffline for both \pTtrig bins. The uncertainties include both statistical and systematic components. The table also shows the 95\% \CL upper limits on \vTwo displayed with the $<$ symbol. In the lower table, \vTwo values for \PhotonP events for $0.3<\pTtrig<3.0\GeV$~\cite{CMSmurillo} in similar \Nch ranges are given.}
	\label{Tab:resultados1}
	\begin{scotch}{l c c c c }
		\multicolumn{5}{c}{$0.3<\pTtrig<3.0\GeV$} \\ 
		\hline
		\Ntrkoffline & $\langle \Nch \rangle$ &$\VoneDelta(\times10^{3})$&$\VtwoDelta(\times10^{3})$&\vTwo upper limit at 95\% \CL \\
		
		\hline
		$[2,5)$ &3.30 & $-68.9\pm2.9$ & $6.3\pm4.8$ & $< 0.12$ \\
		$[5,7)$ &6.60 & $-63.3\pm3.7$ & $6.2\pm3.8$ & $< 0.11$ \\
		$[7,40)$ &10.48 & $-61.5\pm3.1$ & $14\pm11$ & $< 0.18$ \\
		$[2,40)$ &4.06 & $-66.5\pm2.1$ & $7.9\pm5.3$ & $< 0.13$ \\
		\hline
		\multicolumn{5}{c}{$1<\pTtrig<3\GeV$} \\ 
		\hline
		$[2,40)$ & 4.07 & $-204\pm23$ & $62\pm26$ & $< 0.32$ \\
	\end{scotch}
\\[0.4cm]
	\begin{scotch}{l c c}
		\multicolumn{3}{c}{\vTwo results for \PhotonP events for $0.3<\pTtrig<3.0\GeV$~\cite{CMSmurillo}} \\
		\hline
		\Ntrkoffline & $\langle \Nch \rangle$ &\vTwo\\
		\hline
		$[2,5)$ & 3.24 & $0.111\pm0.019$ \\
		$[5,10)$ & 7.15 & $0.123\pm0.018$ \\
		$[10,35)$ & 13.87 & $0.161\pm0.020$ \\
	\end{scotch}
\end{table*}

\clearpage
\section{Summary}
Long-range azimuthal $(\phi)$ two-particle correlations have been studied for in proton-lead (\pPb) events at $\sqrtsNN=8.16\TeV$ as a function of the forward rapidity gap width (\DeltaEtaF) and the charged-particle multiplicity (\Nch), where \pt is the transverse momentum. Varying \DeltaEtaF provides a continuous handle on the degree of exclusivity of the selected \pPb events and therefore complements inclusive small-system correlation measurements. Events have been selected by requiring energy deposition in one of the forward calorimeters and no activity in the forward calorimeter on the opposite side. In addition, we require no energy in the region $2.5 < \eta < 3$, favoring events with activity in the direction of the outgoing Pb beam. For events with $\DeltaEtaF > 2.5$, the fraction of diffractive events is enhanced, and these events were further studied as a function of charged particle multiplicity. The average track multiplicity is lower in the diffractive sample than for inclusive \pPb events, but larger than for photon-proton (\PhotonP) events. The slope of the multiplicity distribution becomes steeper with increasing \DeltaEtaF. The mean track multiplicity $\langle \Ntrkoffline \rangle$ decreases by 48\% when \DeltaEtaF changes from $0.5<\DeltaEtaF<1.0$ to $\DeltaEtaF>2.5$. The two-dimensional two-particle correlation functions have been averaged over $\abs{\DeltaEta}>2$ and the resulting \DeltaPhi distributions have been studied in terms of their Fourier coefficients \VoneDelta and \VtwoDelta, as a function of \DeltaEtaF and \Nch. 

The measured values of \VoneDelta are negative and decrease with \DeltaEtaF, while the central values of \VtwoDelta increase monotonically with \DeltaEtaF, although they are consistent with zero within uncertainties in two of the bins. The results have been compared with the predictions of the \PYTHIA 8.3.10 model, which does not include collective effects, and with \textsc{epos-lhc} model, which includes such effects. The measured values of the elliptic flow coefficient (\vTwo) are provided for the bins $[0.5,1)$, $[1,1.5)$, $[2,2.5)$. In the other bins, the \vTwo results are consistent with zero. In these bins, 95\% confidence level upper limits have been determined; they are stable across the \DeltaEtaF and \Nch regions studied, indicating comparable sensitivity in all bins, and reach values up to 0.13 for the most inclusive \Ntrkoffline bin. The values of the upper limits are similar to the values measured in \PhotonP events.

These results constrain scenarios in which ridge-like correlations are expected to persist universally across small systems independent of event topology. They also provide a reference for models that predict reduced collective response when the multiparton interactions and color reconnection processes are suppressed. 

\begin{acknowledgments}
We congratulate our colleagues in the CERN accelerator departments for the excellent performance of the LHC and thank the technical and administrative staffs at CERN and at other CMS institutes for their contributions to the success of the CMS effort. In addition, we gratefully acknowledge the computing centers and personnel of the Worldwide LHC Computing Grid and other centers for delivering so effectively the computing infrastructure essential to our analyses. Finally, we acknowledge the enduring support for the construction and operation of the LHC, the CMS detector, and the supporting computing infrastructure provided by the following funding agencies: SC (Armenia), BMBWF and FWF (Austria); FNRS and FWO (Belgium); CNPq, CAPES, FAPERJ, FAPERGS, and FAPESP (Brazil); MES and BNSF (Bulgaria); CERN; CAS, MoST, and NSFC (China); MINCIENCIAS (Colombia); MSES and CSF (Croatia); RIF (Cyprus); SENESCYT (Ecuador); ERC PRG and PSG, TARISTU24-TK10 and MoER TK202 (Estonia); Academy of Finland, MEC, and HIP (Finland); CEA and CNRS/IN2P3 (France); SRNSF (Georgia); BMFTR, DFG, and HGF (Germany); GSRI (Greece); MATE and NKFIH (Hungary); DAE and DST (India); IPM (Iran); SFI (Ireland); INFN (Italy); MSIT and NRF (Republic of Korea); MES (Latvia); LMTLT (Lithuania); MOE and UM (Malaysia); BUAP, CINVESTAV, CONACYT, LNS, SEP, and UASLP-FAI (Mexico); MOS (Montenegro); MBIE (New Zealand); PAEC (Pakistan); MSHE, NSC, and NAWA (Poland); FCT (Portugal); MESTD (Serbia); MICIU/AEI and PCTI (Spain); MOSTR (Sri Lanka); Swiss Funding Agencies (Switzerland); MST (Taipei); MHESI (Thailand); TUBITAK and TENMAK (T\"{u}rkiye); NASU (Ukraine); STFC (United Kingdom); DOE and NSF (USA).

\begin{sloppy}
\setlength\emergencystretch{\hsize}
\hyphenation{Rachada-pisek} Individuals have received support from the Marie-Curie program and the European Research Council and Horizon 2020 Grant, contract Nos.\ 675440, 724704, 752730, 758316, 765710, 824093, 101115353, 101002207, 101001205, and COST Action CA16108 (European Union); the Leventis Foundation; the Alfred P.\ Sloan Foundation; the Alexander von Humboldt Foundation; the Science Committee, project no. 22rl-037 (Armenia); the Fonds pour la Formation \`a la Recherche dans l'Industrie et dans l'Agriculture (FRIA) and Fonds voor Wetenschappelijk Onderzoek contract No. 1228724N (Belgium); the Beijing Municipal Science \& Technology Commission, No. Z191100007219010, the Fundamental Research Funds for the Central Universities, the Ministry of Science and Technology of China under Grant No. 2023YFA1605804, the Natural Science Foundation of China under Grant No. 12535004, and USTC Research Funds of the Double First-Class Initiative No.\ YD2030002017 (China); the Ministry of Education, Youth and Sports (MEYS) of the Czech Republic; the Shota Rustaveli National Science Foundation (Georgia); the Deutsche Forschungsgemeinschaft (DFG), among others, under Germany's Excellence Strategy -- EXC 2121 ``Quantum Universe" -- 390833306, and under project number 400140256 - GRK2497; the Hellenic Foundation for Research and Innovation (HFRI), Project Number 2288 (Greece); the Hungarian Academy of Sciences, the New National Excellence Program - \'UNKP, the NKFIH research grants K 131991, K 138136, K 143460, K 143477, K 147557, K 146913, K 146914, K 147048, TKP2021-NKTA-64, and 2025-1.1.5-NEMZ\_KI-2025-00004, and MATE KKP and KKPCs Research Excellence and Flagship Research Groups grants (Hungary); the Council of Science and Industrial Research, India; ICSC -- National Research Center for High Performance Computing, Big Data and Quantum Computing, FAIR -- Future Artificial Intelligence Research, and CUP I53D23001070006 (Mission 4 Component 1), funded by the NextGenerationEU program, the Italian Ministry of University and Research (MUR) under Bando PRIN 2022 -- CUP I53C24002390006, PRIN PRIMULA 2022RBYK7T (Italy); the Latvian Council of Science; the Ministry of Science and Higher Education, project no. 2022/WK/14, and the National Science Center, contracts Opus 2021/41/B/ST2/01369, 2021/43/B/ST2/01552, 2023/49/B/ST2/03273, and the NAWA contract BPN/PPO/2021/1/00011 (Poland); the Funda\c{c}\~ao para a Ci\^encia e a Tecnologia (Portugal); the National Priorities Research Program by Qatar National Research Fund; MICIU/AEI/10.13039/501100011033, ERDF/EU, ``European Union NextGenerationEU/PRTR", projects PID2022-142604OB-C21, PID2022-139519OB-C21, PID2023-147706NB-I00, PID2023-148896NB-I00, PID2023-146983NB-I00, PID2023-147115NB-I00, PID2023-148418NB-C41, PID2023-148418NB-C42, PID2023-148418NB-C43, PID2023-148418NB-C44, PID2024-158190NB-C22, RYC2021-033305-I, RYC2024-048719-I, CNS2023-144781, CNS2024-154769 and Plan de Ciencia, Tecnolog{\'i}a e Innovaci{\'o}n de Asturias, Spain; the Chulalongkorn Academic into Its 2nd Century Project Advancement Project, the National Science, Research and Innovation Fund program IND\_FF\_68\_369\_2300\_097, and the Program Management Unit for Human Resources \& Institutional Development, Research and Innovation, grant B39G680009 (Thailand); the Eric \& Wendy Schmidt Fund for Strategic Innovation through the CERN Next Generation Triggers project under grant agreement number SIF-2023-004; the Kavli Foundation; the Nvidia Corporation; the SuperMicro Corporation; the Welch Foundation, contract C-1845; and the Weston Havens Foundation (USA).
\end{sloppy}
\end{acknowledgments}\section*{Data availability} Release and preservation of data used by the CMS Collaboration as the basis for publications is guided by the  \href{https://doi.org/10.7483/OPENDATA.CMS.1BNU.8V1W}{CMS data preservation, re-use and open access policy}.

\bibliography{auto_generated}

\cleardoublepage \appendix\section{The CMS Collaboration \label{app:collab}}\begin{sloppypar}\hyphenpenalty=5000\widowpenalty=500\clubpenalty=5000\input{HIN-22-004-public-authorlist.tex}\end{sloppypar}
\end{document}

%% file: HIN-22-004-public-authorlist.tex
\cmsinstitute{Yerevan Physics Institute, Yerevan, Armenia}
{\tolerance=6000
A.~Hayrapetyan, A.~Tumasyan\cmsAuthorMark{1}\cmsorcid{0009-0000-0684-6742}
\par}
\cmsinstitute{Institut f\"{u}r Hochenergiephysik, Vienna, Austria}
{\tolerance=6000
W.~Adam\cmsorcid{0000-0001-9099-4341}, J.W.~Andrejkovic, L.~Benato\cmsorcid{0000-0001-5135-7489}, T.~Bergauer\cmsorcid{0000-0002-5786-0293}, S.~Chatterjee\cmsorcid{0000-0003-2660-0349}, K.~Damanakis\cmsorcid{0000-0001-5389-2872}, M.~Dragicevic\cmsorcid{0000-0003-1967-6783}, P.S.~Hussain\cmsorcid{0000-0002-4825-5278}, M.~Jeitler\cmsAuthorMark{2}\cmsorcid{0000-0002-5141-9560}, N.~Krammer\cmsorcid{0000-0002-0548-0985}, A.~Li\cmsorcid{0000-0002-4547-116X}, D.~Liko\cmsorcid{0000-0002-3380-473X}, I.~Mikulec\cmsorcid{0000-0003-0385-2746}, J.~Schieck\cmsAuthorMark{2}\cmsorcid{0000-0002-1058-8093}, R.~Sch\"{o}fbeck\cmsAuthorMark{2}\cmsorcid{0000-0002-2332-8784}, D.~Schwarz\cmsorcid{0000-0002-3821-7331}, M.~Sonawane\cmsorcid{0000-0003-0510-7010}, W.~Waltenberger\cmsorcid{0000-0002-6215-7228}, C.-E.~Wulz\cmsAuthorMark{2}\cmsorcid{0000-0001-9226-5812}
% ...
\par}
\vskip\cmsinstskip
\dag:~Deceased\\
$^{1}$Also at Yerevan State University, Yerevan, Armenia\\
$^{2}$Also at TU Wien, Vienna, Austria\\
% ...
